# Multi-Stream Transmission in Cell-Free MIMO Networks with Coherent AP Clustering

Esra Aycan Beyazıt, Jeroen Famaey, Nina Slamnik-Krije štorac, Johann M. Marquez-Barja, Miguel Camelo Botero
University of Antwerp - imec, IDLab, Antwerp, Belgium
Email: {esra.aycanbeyazit, jeroen.famaey, nina.slamnik-krijestorac,
johann.marquez-barja, miguel.camelobotero}@uantwerpen.be

*Abstract*—This letter proposes a multi-stream selection framework for Cell-Free Multiple-Input Multiple-Output (CF-MIMO) networks. Partially coherent transmission has been considered by clustering Access Points (APs) into phase-aligned clusters to address the challenges of phase misalignment and inter-cluster interference. A novel stream selection algorithm is developed to dynamically allocate multiple streams to each multi-antenna User Equipment (UE), ensuring that the system optimizes the sum rate while minimizing inter-cluster and inter-stream interference. Numerical results validate the effectiveness of the proposed method in enhancing spectral efficiency and fairness in distributed CF-MIMO networks.

## I. Introduction

The deployment of CF-MIMO has emerged as a key enabler for next-generation wireless networks, offering substantial improvements in spectral efficiency, energy efficiency, and uniform user coverage by eliminating cell boundaries [1]. In CF-MIMO networks, a large number of geographically distributed APs collaboratively serve UEs on the same time-frequency resources, achieving high data rates through coherent transmission [2]. However, practical limitations such as hardware imperfections, synchronization errors, and varying propagation delays introduce phase misalignment among APs, significantly impacting the effectiveness of coherent operation. Specifically, these factors hinder accurate Channel State Information (CSI) acquisition and disrupt phase alignment, leading to severe degradation in the achievable rate of the network.

To address the challenges of phase misalignment and asynchronous reception, recent studies have proposed a mixed coherent and non-coherent transmission approach [3]. Since clustering APs within a small area enables coherent transmission [2], APs are grouped into phase-aligned clusters, where those within the same cluster perform coherent transmission, while transmission across clusters remains non-coherent. This hybrid strategy mitigates performance degradation caused by asynchronous reception by balancing the benefits of coherent transmission within clusters with the flexibility of non-coherent transmission between them. In the typical case where each UE is equipped with a single antenna, it receives the same data stream from all APs within its assigned cluster, whereas APs in different clusters transmit distinct data streams.

Most state-of-the-art solutions [4]–[6] assume UEs are equipped with a single antenna in CF-MIMO networks, which limits their applicability in next-generation networks, where multi-antenna UEs are becoming increasingly common. When UEs are equipped with multiple antennas and multiple data streams are transmitted simultaneously, interference between these streams can significantly degrade overall system performance, especially in networks with mixed coherent and non-coherent transmission regimes [3]. Consequently, managing inter-stream interference in non-coherent regimes is crucial to ensure that multi-stream transmissions provide a significant performance boost compared to single-stream transmissions [7]. Stream selection based on Interference Alignment (IA) was initially studied in conventional and heterogeneous cellular networks [8], [9]. However, despite its importance for enhancing spectral efficiency, determining the optimal number of streams for each UE and selecting the best stream combinations in CF-MIMO networks remains underexplored.

To address these challenges, this paper investigates multi-stream selection in CF-MIMO networks with mixed coherent and non-coherent transmission. We first analyze the impact of multi-stream transmission in partially coherent CF-MIMO systems, examining how clustering and transmission strategies affect sum-rate and inter-cluster interference. To mitigate interference and improve fairness, we propose a dynamic stream allocation algorithm that optimally assigns data streams to UEs equipped with multiple antennas. The approach jointly computes precoding and decoding matrices to maximize spectral efficiency while ensuring each UE receives at least one stream. Stream sequences are initialized with the strongest streams from each cluster-UE pair and compared to determine the sequence achieving the highest sum rate. The framework dynamically adjusts the number of streams per UE based on network conditions, balancing throughput and interference mitigation. The proposed algorithm is evaluated through extensive numerical simulations under diverse clustering scenarios, comparing it to both an upper bound and a baseline. The upper bound is obtained via an exhaustive search of all possible stream combinations, identifying the optimal sequence. The baseline follows a greedy approach, constructing a single stream sequence by iteratively selecting the strongest streams for each UE to maximize sum-rate. Our method demonstrates significant improvements in sum-rate, particularly in scenarios with severe phase misalignment and interference.

## II. System Model

This study considers a CF-MIMO system with $L$ APs, each with $N_T$ transmit antennas, serving $K$ UEs, each with $N_R$

receive antennas. All APs are connected to a central processing unit (CPU) via fronthaul links. While full phase coherence is ideal, network-wide synchronization is impractical [2]. Thus, the network supports both coherent and non-coherent transmissions: APs within a cluster maintain phase alignment for coherent transmission, whereas inter-cluster transmissions remain non-coherent due to the lack of synchronization.

The system model for both transmission modes is defined as follows: Let $\mathcal{M}_k \subset \{1, \ldots, L\}$ denote the subset of APs serving UE $k$, referred to as clusters throughout this paper. A fully non-coherent transmission occurs when $|\mathcal{M}_k| = L$, meaning no clustering exists. Conversely, when $|\mathcal{M}_k| = 1$, a single large cluster enables fully coherent transmission.

### A. Downlink Data Transmission

The considered phase-aligned transmission inside the clusters is achieved by forming a virtual large MIMO array, enabling the coherent transmission of data symbols [10]. Each cluster is associated with one UE. The set of UE-cluster pairs can be denoted as $k \in \Gamma = \{1, \ldots, K\}$. The implemented coherent AP clustering method is described in Section III-A.

If the $c^{th}$ cluster is denoted by $\mathcal{M}_c$, in a coherent transmission, the channel between the subset of APs $\mathcal{M}_c$ and UE $k$ is represented by the collective channel $\mathbf{H}_{k\mathcal{M}_c} \in \mathbb{C}^{N_R \times N_T |\mathcal{M}_c|}$, where $|\mathcal{M}_c|$ is the number of APs in cluster $\mathcal{M}_c$. The collective channel between cluster $\mathcal{M}_c$ and UE $k$ can be expressed as:

$$\mathbf{H}_{k\mathcal{M}_c} = \begin{bmatrix} \mathbf{h}_{kl_1} & \mathbf{h}_{kl_2} & \cdots & \mathbf{h}_{kl_{|\mathcal{M}_c|}} \end{bmatrix}, \quad (1)$$

where $l_1, l_2, \ldots, l_{|\mathcal{M}_c|} \in \mathcal{M}_c$ and $\mathbf{h}_{kl} \in \mathbb{C}^{N_R \times N_T}$ represents the channel matrix from AP $l$ to UE $k$, for all $l \in \mathcal{M}_c$.

The output signal at user $k$ is defined as follows.

$$\mathbf{y}_k = \alpha_{kk} \mathbf{H}_{k\mathcal{M}_k} \mathbf{x}_k + \sum_{\substack{j=1, \\ j \neq k}}^{K} \alpha_{kj} \mathbf{H}_{k\mathcal{M}_j} \mathbf{x}_j + \mathbf{n}_k \quad (2)$$

where, $\alpha_{kj} \mathbf{H}_{k\mathcal{M}_j}$ is the channel matrix between cluster $\mathcal{M}_j$ and UE $k$ with dimension $N_R \times N_T |\mathcal{M}_j|$. Each element of $\mathbf{H}_{k\mathcal{M}_j}$ includes fading, which is modeled as an independent and identically distributed complex Gaussian random variable with $\mathcal{CN}(0,1)$. $\alpha_{kj}$ is the large-scale fading coefficient and it can be modeled as [1].

$$\alpha_{kl} = 10^{-\text{PL}(\text{dis}_{kl})/10} 10^{-F_{kl}/10}, \quad (3)$$

where $\text{PL}(\text{dis}_{kl})$ represents the path loss function with the parameter of the distance between AP $l$ and UE $k$, and $F_{kl}$ is the shadowing effect. For each receiver $k$, $\mathbf{n}_k$ is a $N_R \times 1$ vector. Each element of $\mathbf{n}_k$ represents additive white Gaussian noise with zero mean and variance of $\sigma^2$. $\mathbf{x}_{\mathcal{M}_k}$ is the transmitted signal from the $\mathcal{M}_k^{th}$ cluster with dimension $N_T |\mathcal{M}_k| \times 1$ and it is calculated as follows.

$$\mathbf{x}_{\mathcal{M}_k} = \sqrt{P_k} \mathbf{T}_{\mathcal{M}_k} \mathbf{s}_k \quad (4)$$

where $P_k$ is the transmit power of AP $k$. $\mathbf{T}_{\mathcal{M}_k}$ is the unitary precoding matrix of cluster $\mathcal{M}_k$ with dimension $N_T |\mathcal{M}_k| \times q_k$, and cluster $\mathcal{M}_k$ can transmit $q_k$ independent streams with $q_k \leq d_k$ where $d_k = \min(N_R, N_T |\mathcal{M}_k|)$. $\mathbf{s}_k$ is the symbol vector with dimension of $q_k \times 1$ and denoted as $\mathbf{s}_k = [s_{k,1} \ldots s_{k,q_k}]^T$ where $\mathbb{E}\left[\|\mathbf{s}_k\|^2\right] = 1$, and it is assumed that the transmit power is equally shared between the symbols, $\mathbb{E}\left[|s_{k,n}|^2\right] = 1/q_k$, $n = 1, \ldots, q_k$. In addition, the maximum total number of streams in the network is calculated as $r = \sum_{k=1}^{K} d_k$. Desired signals are obtained by multiplying $\mathbf{y}_k$ with the postcoding vector, $\mathbf{D}_k$ with a size of $N_R \times q_k$. The obtained decoded data symbols can be written as

$$\hat{\mathbf{y}}_k = \mathbf{D}_k^H \mathbf{y}_k \quad (5)$$

The data rate for the $i^{th}$ stream of the $k^{th}$ user can be expressed as follows.

$$\mathrm{R}_{ki} = \log_2(1 + \gamma_{ki}) \quad (6)$$

where $\gamma_{ki}$ is the SINR for the $i^{th}$ stream of the $k^{th}$ user and it is calculated as

$$\gamma_{ki} = \frac{(P_k/q_k)\alpha_{kk}^2 \mathbf{d}_k^{iH} \mathbf{H}_{k\mathcal{M}_k} \mathbf{t}_k^i \mathbf{t}_k^{iH} \mathbf{H}_{k\mathcal{M}_k}^H \mathbf{d}_k^i}{\mathbf{d}_k^{iH} \mathbf{B}_{k\mathcal{M}_i} \mathbf{d}_k^i} \quad (7)$$
$$\forall k = 1, \ldots, K, \quad \forall i = 1, \ldots, q_k$$

where $\mathbf{t}_k^i$ is the $i^{th}$ column vector of the precoding matrix $\mathbf{T}_k$ with dimension $N_T |\mathcal{M}_k| \times 1$, and $\mathbf{d}_k^i$ is the $i^{th}$ column vector of postcoding matrix $\mathbf{D}_k$ with dimension $N_R \times 1$. Furthermore, $\mathbf{B}_{ki}$ is defined as the interference plus noise covariance matrix for the $i^{th}$ stream of the $k^{th}$ receiver and it is given by

$$\mathbf{B}_{ki} = \sum_{\substack{l=1, \\ l \neq i}}^{q_k} \frac{P_k}{q_k} \alpha_{kk}^2 \mathbf{H}_{k\mathcal{M}_k} \mathbf{t}_k^l (\mathbf{t}_k^l)^H \mathbf{H}_{k\mathcal{M}_k}^H + \quad (8)$$
$$\sum_{\substack{j=1 \\ j \neq k}}^{K} \sum_{q=1}^{q_j} \frac{P_j}{q_j} \alpha_{kj}^2 \mathbf{H}_{k\mathcal{M}_j} \mathbf{t}_j^q (\mathbf{t}_j^q)^H \mathbf{H}_{k\mathcal{M}_j}^H + \sigma^2 \mathbf{I}_{N_R},$$
$$\forall k = 1, \ldots, K, \quad \forall i = 1, \ldots, q_k.$$

Accordingly, the sum rate (SR) is calculated as follows.

$$\mathrm{SR} = \sum_{k=1}^{K} \sum_{i=1}^{q_k} \log_2(1 + \gamma_{ki}) \quad (9)$$

### B. Problem Definition

The main objective is to minimize interference while identifying the optimal stream allocation scheme for each AP cluster and UE within the system. In this context, the stream allocation problem aims to maximize the total sum rate of the network while ensuring that each user has at least one stream selected, thus guaranteeing service. Mathematically, this can be formulated as follows.

$$\left\{ (\mathbf{T}_{\mathcal{M}_k}^*, \mathbf{D}_k^*) \right\}_{k=1,\ldots,K} = \underset{\mathbf{T}_{\mathcal{M}_k}, \mathbf{D}_k}{\operatorname{argmax}} \mathrm{SR} \quad (10a)$$

$$s.t. \quad d_k \geq 1 \quad k = 1, \ldots, K \quad (10b)$$

where $d_k$ is the number of assigned streams for user $k$.

To mitigate phase misalignment in CF-MIMO networks, existing approaches group phase-coherent APs into clusters [3], [7]. However, inter-cluster interference remains due to the lack of phase alignment across clusters. Additionally, determining the optimal number of streams per UE to maximize throughput while minimizing both inter-cluster and intra-stream interference remains an open challenge. To bridge this gap, we propose a stream selection algorithm that achieves near-optimal performance comparable to exhaustive search methods but with significantly lower computational complexity.

## III. THE PROPOSED FRAMEWORK

Motivated by the existing gap in the literature, this section presents a stream selection algorithm for a coherently clustered network aimed at optimizing the sum-rate and minimizing interference in a CF-MIMO network. The proposed framework mainly consists of AP clustering and stream selection. First, APs are grouped into clusters based on proximity and reference distance which was previously proposed in the studies of [3], [7]. Next, system parameters are initialized, and CSI is computed. The key contribution of this work lies in the stream selection step, where inter-cluster interference is managed through orthogonal projections after each selection, effectively suppressing interference both to and from the selected stream. The overall framework is summarized in Algorithm 1.

### A. AP Clustering for Coherent Transmission

This sub-section explains a clustering algorithm that enables mixed coherent and non-coherent transmissions in a distributed CF-MIMO network by grouping APs based on their proximity to UEs and a reference distance $D_{\text{ref}}$ [3], [7]. Achieving a full phase coherency among geographically distributed APs is practically challenging due to independent local oscillators and varying propagation delays, which result in phase misalignment across the network. The proposed approach mitigates these phase mismatches by ensuring phase-aligned APs within each cluster, while inter-cluster transmissions remain non-coherent. This design balances the trade-off between beamforming gain and network scalability, making it more feasible for large-scale deployments.

For a user $k$, the algorithm starts by selecting the closest AP, denoted as $l_k$, and forms a cluster $\mathcal{M}_1$ by including all neighboring APs $l'$ satisfying $D_{ll'} \leq D_{\text{ref}}$, where $D_{ll'}$ is the distance between APs $l$ and $l'$ [2]. Assigned APs are removed from the available set, and the process iterates by selecting the next closest unassigned AP $l_n$ to form $\mathcal{M}_n$. This continues until all APs are assigned to distinct, non-overlapping clusters, $\mathcal{M} = \{\mathcal{M}_1, \mathcal{M}_2, \ldots, \mathcal{M}_{L_c}\}$, where $\mathcal{M}$ is the complete set of clusters and $L_c$ is the total number of formed clusters. Further details are available in [7]. In our work, each UE is associated with the cluster that provides the strongest channel gain. Inter-cluster interference is addressed through precoding and combining techniques discussed in Section III-B.

### B. Interference Mitigation

In stream selection-based IA algorithms, each stream is chosen to lie in the null space of previously selected streams, ensuring interference avoidance. The null space refers to vectors that, when multiplied by the channel matrix, yield zero interference.

**Alg. 1** The Proposed Framework Flow

**Input:** Set of APs, UEs, CSIs.
**Output:** Formed AP clusters, selected streams, and corresponding beamforming vectors.
Coherent AP Clustering by implementing **Algorithm 1 in [7]**.
Initialize system parameters and compute CSI.
Call **Algorithm 2** to perform the proposed stream selection algorithm based on the formed clusters.

Streams are computed using the Singular Value Decomposition (SVD) of all channels, $(\alpha_{kk}\mathbf{H}_{k\mathcal{M}_k}) = \mathbf{U}_k \mathbf{S}_k \mathbf{V}_{\mathcal{M}_k}^H$ where $\mathbf{U}_k$ and $\mathbf{V}_{\mathcal{M}_k}$ are orthogonal matrices representing receive and transmit beamforming directions at the UE and APs, respectively, and $\mathbf{S}_k$ contains the singular values indicating stream strengths. The $l^{th}$ column vectors of $\mathbf{U}_k$ and $\mathbf{V}_{\mathcal{M}_k}$ are denoted by $\mathbf{u}_k^l$ and $\mathbf{v}_{\mathcal{M}_k}^l$.

To mitigate interference, IA techniques align the interfering components after each stream selection step. Two interference types are considered: (i) from the selected stream to remaining streams, and (ii) from remaining streams to the selected stream. Correspondingly, two virtual channels, Virtual Receiving Channel (VRC) and Virtual Transmitting Channel (VTC), are defined [8]. Precoding and postcoding matrices, constructed from the selected stream vectors, are expressed as: $\mathbf{T}_{\mathcal{M}_k^*} = [\mathbf{v}_{\mathcal{M}_k^*}^1, \ldots, \mathbf{v}_{\mathcal{M}_k^*}^{q_k}]$ and $\mathbf{D}_{k^*} = [\mathbf{u}_{k^*}^1, \ldots, \mathbf{u}_{k^*}^{q_k}]$. Then, interference is mitigated through orthogonal projections. For users $j \neq k$, the remaining beamformers are projected onto the null space of the selected stream's VRC and VTC, yielding projected matrices $\mathbf{H}_{j\mathcal{M}_j}^{\perp}$. At iteration $i$, interference from and to the selected stream is reduced by projecting channel matrices orthogonally to the respective virtual channels. The projection matrix is given by $\mathbf{P}_{\mathbf{x}}^{\perp} = \mathbf{I} - \frac{\mathbf{x}\mathbf{x}^H}{\|\mathbf{x}\|^2}$. The complete IA procedure is detailed in Algorithm 1 of [9].

### C. Multi-Stream Transmission in CF-Networks

In this section, we propose a recursive stream selection procedure to determine the optimal beam combinations while incorporating the IA approach explained in sub-section III-B at each stream selection step. The process initializes multiple stream sequences, each beginning with the strongest stream for every cluster-user pair, defined as the one with the highest singular value of the channel matrix. The number of initialized sequences equals the number of cluster-user pairs, with the initial set, $\Omega_0$, containing only the best streams of each pair.

After initializing the stream paths, the selection procedure recursively chooses the next stream that maximizes the sum-rate at each iteration. This stream is selected from the set $\Omega$, which tracks all available streams. Suppose no stream can be selected to increase the sum-rate during iteration $i$. In that case, the algorithm selects the stream that causes the minimum sum-rate decrease, prioritizing users with no previously selected streams. The process continues until no further streams can be selected. The selected streams are stored in a set denoted as $\Psi$. At each iteration of the proposed *Comparative Stream Selection (CSS)* algorithm, the sets $\Omega$ and $\Psi$ are updated: the chosen stream is removed from $\Omega$ and added to $\Psi$. The whole process is given in Algorithm 2.

**Alg. 2** CSS Algorithm
---
Construct the initialization set $\Omega_0$
  $\Omega_0 = \{(k,l) \mid k \in \Gamma \text{ and } l = 1\}$
Start constructing stream sequences
**for** each stream $(k^*, l^*) \in \Omega_0$ **do**
  Initialize the variables
    $\Psi = \emptyset;\ i = 1;\ d_k = 0;\ \text{finish} = \text{FALSE}$ and
    $\mathbf{H}^{\perp}_{k_k \mathcal{M}_k} = \mathbf{H}_{k \mathcal{M}_k}$ for $k = 1, ..., K$
  Compute the SVD of all couples
    $\mathbf{H}_{k \mathcal{M}_k} = \mathbf{U}_k \mathbf{S}_k \mathbf{V}^H_{\mathcal{M}_k}$ for $k = 1, ..., K$
  Set the stream to be selected initially $(k^*, l^*)$
    $\Psi = \Psi \cup (k^*, l^*)$
    $d_{k^*} = d_{k^*} + 1$
  **Perform Algorithm 1 in [9]**.
  Construct $\Omega = \left\{ (\mathbf{S}_k)(l,l) \mid k = 1, ..., K \text{ and } l = 1, ..., \mathbf{rank}(\mathbf{H}^{\perp}_{k \mathcal{M}_k}) \right\}$
  Continue selecting streams by applying **Algorithm 2 in [9]**
  Compute $(\mathbf{T}_k)_\Psi$, $(\mathbf{D}_k)_\Psi$ and $SR_\Psi$ for stream sequence $\Psi$
**end for**
Select the best stream sequence according to Eq. (10a)
  $\Psi^* = \underset{\Psi}{\arg\max}\ SR_\Psi$
  $\mathbf{T}^*_k = (\mathbf{T}_k)_{\Psi^*},\ \mathbf{D}^*_k = (\mathbf{D}_k)_{\Psi^*}$ for $k = 1, ..., K$
Output: $\mathbf{T}^*_k, \mathbf{D}^*_k\ \forall k$

## IV. Performance Results

In this section, the performance of the proposed algorithm, CSS, is evaluated in a CF-MIMO system setup where users are independently and uniformly distributed in a $1 \times 1$km square area with a wrap-around topology. The minimum distance between each AP is 50m and $D_{ref} = 200$m. System parameters used in the simulations are listed in Table I.

To analyze the performance of the proposed algorithm in CF-MIMO networks, an upper bound is obtained by implementing an exhaustive stream selection algorithm that searches all possible stream sequences. The most challenging drawback of the exhaustive search is the complexity, which depends on the number of streams. In terms of the invoking number of Algorithm 1 in the study of [9] in each stream selection, the complexity of the exhaustive search can be formulated as follows.

$$\sum_{i=K}^{r} \left( \underbrace{\left( i! \left[ \prod_{k=1}^{K} \binom{q_k}{1} \right] \binom{r-K}{i-K} \right)}_{\substack{\text{The total number of} \\ \text{stream sequences of length } i}} \times \underbrace{i}_{\substack{\text{The number of invoking Alg.1 in [9]} \\ \text{for each stream sequence}}} \right).$$

Meanwhile, the complexity of the CSS algorithm can be expressed as $K \times r$, representing a significant reduction in computational complexity achieved by the proposed algorithm.

Due to the high complexity of exhaustive search, we evaluated the upper bound only for a small-scale scenario ($L = 6$, $K = 4$), as shown in Figure 1a. At CDF $= 0.9$, the CSS algorithm achieves a sum-rate of 36 bps/Hz, reaching 86% of the exhaustive search performance, demonstrating its near-optimal efficiency with significantly lower complexity. Additionally, CSS outperforms the greedy approach by 16%, highlighting the benefits of adaptive stream selection. It also surpasses fixed stream allocation schemes, emphasizing the importance of dynamically adjusting streams to network conditions. Notably, allocating all streams when $d = 2$ results in excessive interference, reinforcing the need for efficient stream selection.

For the larger-scale scenario where $L = 24$ and $K = 6$, the CSS algorithm demonstrates once again a strong performance, as illustrated in Figure 1b. At CDF $= 0.9$, CSS achieves a sum-rate of 39 bps/Hz, outperforming the greedy approach by 11%. Similarly, the fixed resolution scheme remains far from optimal. Notably, among all computationally feasible algorithms, CSS achieves the highest sum-rate performance, making it the most efficient practical solution for this scenario.

The complexity comparisons is given in Table II in terms of the number of calls to Algorithm 1 in [9].

Moreover, Figure 2 shows the CDF of the sum-rate for various stream selection approaches, analyzing the impact of interference mitigation and clustering strategies for $L = 12$ and $K = 4$. The results highlight significant differences between the proposed CSS algorithm and the greedy stream selection approaches, both with and without IA, in coherent and non-coherent clustering scenarios. At CDF $= 0.9$, the CSS algorithm with IA in coherent clusters achieves a sum-

**TABLE I:** System Parameters

| Parameter Name | Parameter Value |
|---|---|
| Transmit Power of APs | 30dBm |
| Bandwidth | 50MHz |
| Noise Power | $-174$dBm/Hz |
| Noise Figure | 7dB |
| Simulation Area | $1 \times 1$km |
| Path loss | $-30.5 - 36.7 \log_{10}\left(\frac{\text{dis}_{kl}}{1\,\text{m}}\right)$dB |
| Shadowing std. dev. | 4dB |
| Antenna number of each AP | 4 |
| Antenna number of each UE | 2 |

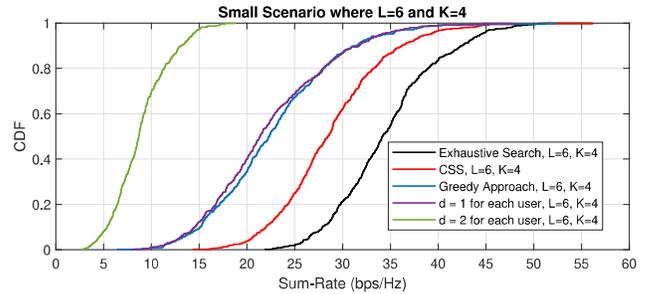

(a) Performance Comparisons in a small scenario where $L = 6$ and $K = 4$

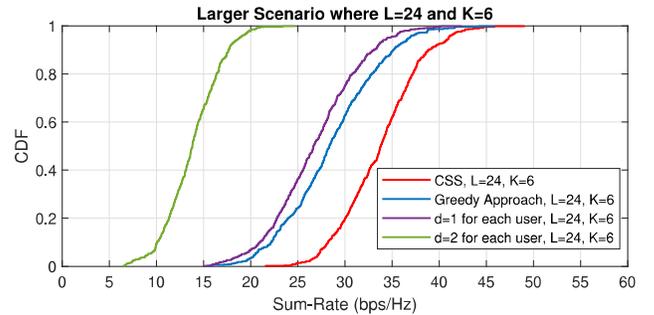

(b) Performance Comparisons in a larger scenario where $L = 24$ and $K = 6$

**Fig. 1:** Performance Comparison of the proposed algorithm in different scales of scenarios

TABLE II: Complexity Comparisons of different stream selection algorithms in terms of invoking Algorithm 1 in [9]

| Algorithm | Small-scale scenario | Large-scale scenario |
|---|---|---|
| Greedy | 2 | 2 |
| CSS | 4 | 6 |
| Exhaustive | $4.9 \times 10^5$ | $\approx 9 \times 10^9$ |

TABLE III: Average number of selected streams and sum-rate values for CSS and Greedy Search across different $N_R$ values.

| $N_R$ | Algorithm | Avg. Number of Streams | Avg. Sum-Rate (bps/Hz) |
|---|---|---|---|
| 2 | CSS Algorithm | 4.05 | 28.84 |
|   | Greedy Search | 4.29 | 22.8 |
| 3 | CSS Algorithm | 5.18 | 31.96 |
|   | Greedy Search | 5.64 | 25.75 |
| 4 | CSS Algorithm | 6.3 | 35.11 |
|   | Greedy Search | 6.69 | 28.10 |

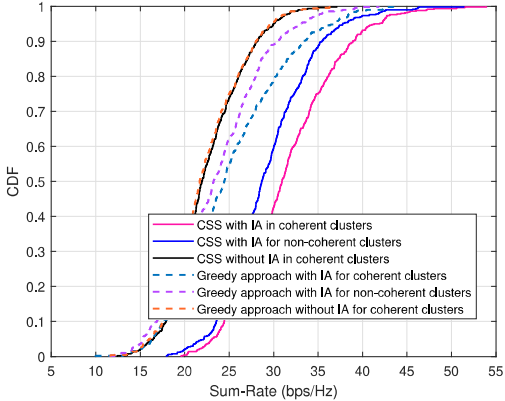

Fig. 2: Impact of coherent clustering and interference mitigation under different schemes on CSS and the greedy approach.

rate of 39 bps/Hz, outperforming the greedy algorithm with IA in coherent clusters by 18%. Similarly, the CSS algorithm with IA in non-coherent clusters achieves 35 bps/Hz, which is 17% higher than the greedy approach in non-coherent clusters. Furthermore, CSS with IA in coherent clusters provides a 11% improvement over CSS with IA in non-coherent clusters. Lastly, CSS with IA in coherent clusters outperforms both CSS and greedy algorithms without IA by 44%, showing the impact of interference mitigation on improving sum-rate performance.

The average number of selected streams and the corresponding average sum-rate values in the small-scale CF-MIMO scenario, where $L = 6$ and $K = 4$, are compared for different numbers of receive antennas at each UE ($N_R = 2, 3, 4$) in Table III. The results show that increasing $N_R$ enables more spatially multiplexed streams, directly improving the sum-rate and spectral efficiency.

## V. CONCLUSION

In this letter, a novel stream selection framework has been proposed for CF-MIMO networks operating under mixed coherent and non-coherent transmission scenarios.

The framework effectively addresses the dual challenges of phase misalignment and inter-cluster interference through dynamic stream allocation and interference mitigation techniques. Simulation results demonstrate the effectiveness of the proposed CSS algorithm, achieving significant improvements in spectral efficiency compared to traditional methods for both small and large-scale networks with severe interference and phase misalignment.

By guaranteeing service fairness while selecting an optimum number of streams for each cluster and UE pairs and reducing computational complexity, the proposed framework is suitable for practical deployment of CF-MIMO systems in next-generation wireless networks. Future work could explore the integration of sub-THz communication and machine learning-based advanced clustering strategies, where a UE can receive streams from other clusters simultaneously to further enhance system performance in ultra-dense networks.

ACKNOWLEDGMENT

This research has been funded by the 6G-TWIN project, which has received funding from the Smart Networks and Services Joint Undertaking (SNS JU) under the EU's Horizon Europe research and innovation program under Grant Agreement No 101136314.